# Goal Directed Relative Skyline Queries in Time Dependent Road Networks

K.B. Priya Iyer[1]

[1]Research Scholar, Sathyabama University

priya_balu_2002@yahoo.co.in

Dr. V. Shanthi[2]

[2]Professor, St. Joseph College of Engineering

drvshanthi@yahoo.co.in

## ABSTRACT

*The Wireless GIS technology is progressing rapidly in the area of mobile communications. Location-based spatial queries are becoming an integral part of many new mobile applications. The Skyline queries are latest apps under Location-based services. In this paper we introduce Goal Directed Relative Skyline queries on Time dependent (GD-RST) road networks. The algorithm uses travel time as a metric in finding the data object by considering multiple query points (multi-source skyline) relative to user location and in the user direction of travelling. We design an efficient algorithm based on Filter phase, Heap phase and Refine Skyline phases. At the end, we propose a dynamic skyline caching (DSC) mechanism which helps to reduce the computation cost for future skyline queries. The experimental evaluation reflects the performance of GD-RST algorithm over the traditional branch and bound algorithm for skyline queries in real road networks.*

## Keywords

*GD-RST, Spatial Databases, Skyline Queries, GIS, Location based services, GPS.*

## 1.  INTRODUCTION

The Wireless technologies are popular in providing location based services to users no matter where they are. The face of Location based services are changing with applications like Indoor Location Positioning, Location-Based Notifications, Social Route Planning, Location-Based Emergency Phone Calls, Navigation & Mapping, Location-Based Advertising etc. A *skyline query* [4] returns all the data objects which are not *dominated* by other objects in data set *D*. For example, a tourist may wish to search for hotel which is cheapest in price, closest to the beach and a hospital. The algorithm now determines the set of all hotels that might be the best choice considering all possible ratios and thus, all possible user preferences.

With the advances in spatial databases, the previous research focuses on the efficiency issues of searching skylines, including block nested loop (BNL) [2], divide- and-conquer (D&C) [2], bitmap and index [23], nearest neighbor (NN) [14], and branch-and-bound (BBS) [18]. It is the first effort to focus on travel time for relative skyline query processing.

To sum up we make the following contributions:

1. Finds the GD-RST by taking travel time as metric.
2. Techniques for reducing the search space and avoid repeated access of data points while finding out relative multi-source nearest neighbor skyline.
3. Designing efficient algorithm in three phases namely Filter, Heap and Refine Skyline phases.

4. Developing cache-centered algorithm to reduce the computation cost of future spatial queries.

The reminder of this paper is organized as follows. In section 2, we review the related work on skyline queries. In section 3, we formally defined GD-RST query in Road Networks. In section 4, we introduce algorithm for finding the data points by considering travel time in road networks. Section 5 presents the results of our experimental evaluation of our proposed approaches with a variety of Spatial Network with large number of data and query objects. Finally section 6 concludes the paper with future research

## 2. RELATED WORK

### 2.1 Single-source Skyline query in Euclidean space

Skyline query processing has been studied extensively in recent years [4, 5, 25, 21, 17, 19, 24]. The *skyline* operator was first introduced into the database community by Borzsonyi et al. [2]. Borzsonyi et al. [2] propose the Block-Nested-Loops algorithm (BNL) and the Extended Divided-and-Conquer algorithm (DC). Both algorithms processes the entire object set for retrieving the skyline data. In [3], the Sort-Filter-Skyline algorithm (SFS) progressively report skyline points by pre-sorting the entire dataset according a preference function. Tan et al.[18] propose a bitmap-based method which transforms each object to bit vectors. Bitmaps can be very large for large values. This method cannot guarantee a good initial response time. Kossmann et al. [7] propose an online nearest neighbor skyline query processing method which can progressively report the skyline points in an order according to user's preference. The objects in the dominated subspace are pruned, and the objects in each non-determined subspace form a to-do list. To remedy this problem, Papadias et al. [14] propose an R-tree based algorithm, Brand and Bound Skyline (BBS), which retrieves skyline points by browsing the R-tree based on the best-first strategy. BBS only visits the intermediate nodes not dominated by any determined skyline points. This method has more efficient memory consumption than the method in [7]. Another study by Lin et al. [11] process a skyline point query against the most recent N elements in a data stream.

### 2.2 Skyline variants

There are many variants of the traditional skyline query. Pei et al. [16] and Yuan et al. [22] proposed methods to compute skylines in all possible subspaces. Tao et al. [19] gave an efficient algorithm to calculate skylines in a specific subspace. Dellis and Seeger [4] proposed a reverse skyline query, which obtains those objects that have the query point as skyline, where each attribute is defined as the absolute difference from objects to query point along each dimension. In the context of uncertain databases, Pei et al. [15] proposed the probabilistic skyline over uncertain data, which returns a number of objects that are expected to be skylines with probability higher than a threshold. The most relevant problems to our work are the dynamic skyline[14], spatial skyline [17], multi-source skyline on road networks [5], multi-preference path planning approach[9] and continuous probabilistic skyline queries over uncertain data streams[8]. Specifically, Papadias et al. [14] applies BBS algorithm to retrieve skyline points, where dynamic attributes of data objects are computed by a set of dimension functions. However, only Euclidean distance was considered for dimension functions. Similarly, the method proposed for multi-source skyline on road networks [5] also utilizes geometric information of data objects during the pruning, which is thus limited to road network application.

In summary, previous studies on skyline variants are limited to either Euclidean space or metric space for a specific application. In contrast, our work focuses on the travel time as a metric for skyline search which play a vital role in road networks. It also helps future queries computation time by caching technique.

## 3. SYSTEM MODEL

In this section, we describe the road network and system model; define the GD-RST query search in spatial networks. We assume a spatial network [California Road Network], containing set of static data objects as well as query objects searching their skyline objects. We assume all road maps and traffic data are maintained by cloud server.

### 3.1  Road Network

We model the underlying road network as a weighted undirected graph G = (V,E) where E is an Edge set of road segments in the road network, V is the Vertex set of intersection points of the road segment and each edge is given travel time of its corresponding road segment as weights.

Figure 1. GD-RST Architecture

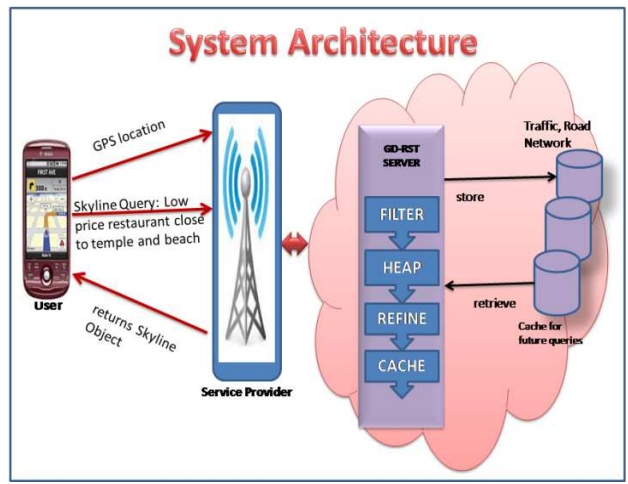

In this model (Figure 1), we consider our system with a mobile environment in which mobile user is able to communicate with the service provider through wireless communication infrastructure e.g.: Wi-Fi.

### 3.2 Definition

A GD-RST query is a Goal Directed Relative Skyline queries on Time dependent (GD-RST) road networks where data objects are returned in the user way of travel relative to their location. The GD-RST query is useful in following situations:

*Example1 :( Query Q1)* if a user wants to find a serviced apartment close to a hospital and a restaurant. From the Fig: 1, the algorithm returns serviced apartment (A3) object which is close to hospital (H4) and restaurant (R1) based on the travel time.

*Example2 :( Query Q2)* if a tourist wants to find a low price restaurant nearer to a temple and close to a beach. From the Fig: 1, the algorithm returns restaurant (R1) object which is nearer to temple (T3) and close to beach (B1) based on the travel time.

## 4. ALGORITHM

### 4.1  Pre-computation

As a pre-computation, the road network is partitioned into grids (Figure 2) to reduce the search space in finding out the nearest neighbor vertex of the user location. We apply a road network clustering approach to efficiently compute the nearest vertex of the query origin and reduce the search space by considering only nodes in user direction of travel.

Figure2. Grid partition – Road Network

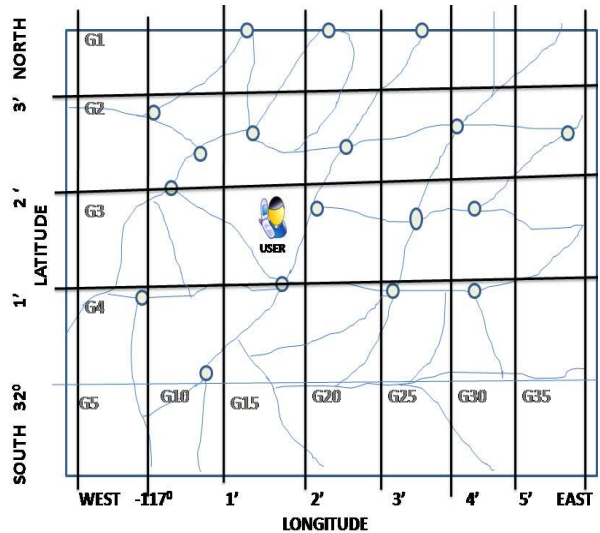

The algorithm first finds the nearest vertex of the query origin. The user location is given by GPS / Wi-Fi. The NEARESTNEIGHBOR NODE function gives the nearest neighboring road node "u" of the user. The road network "R" is divided into $G_1, G_2, G_3....G_n$ grids by taking Latitude, Longitude as grid size. This is done by the function MapGrid. It maps each node in road to a grid. The advantage in partitioning is reducing the search space and increasing the computation speed. Harvesian Formula is used in finding the distance between two locations. The nodes within grid $G_j$ are sorted in descending order of their distances from user location "u" where the top most nodes gives the nearest neighbour of "u".

The algorithm consists of Filter, Heap and Skyline-refine phases. During the Filter phase (refer Figure 3), instead of searching the entire database for list of restaurants, user preferred restaurants that are budgetary within user acceptable range are applied. The resultant candidate set is input to the Heap phase. During Heap phase, a certain degree of aggregation is required for factors such as hotel rating, travel time, total price etc. During this phase, the network is expanded which explores the graph based on travel time to identify other query point nearest neighbor nodes. At the end, for all list of candidate sets, Skyline-refine phase is applied to liter out dominated objects that cannot be in the query answer. At this phase, a caching method is applied which stores the query result for future queries.

From the nearest node, all adjacent nodes are traversed to find the list of first user preferred query point. While searching the query points, the query filters out the unwanted data points that are out of range of user preferences (Filter phase). The algorithm proceeds by applying the other query preferences on the list of candidate data objects (Heap phase). All the candidate objects are refined by considering the minimum travel time and other factors (Skyline-Refine phase). At the end, the algorithm returns the goal directed skyline object relative to user location.

Figure 3. Road Network : GD-RST Query

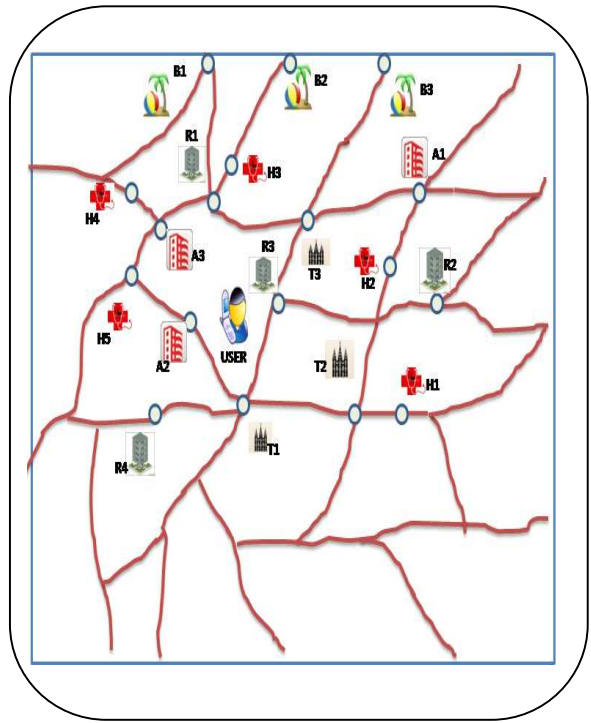

(in above Figure3: R1,R2,R3 – restaurants, H1,H2,H3 – hospitals, T1,T2,T3 – temples, B1,B2,B3 – beach).

## 4.2    Algorithm

## Algorithm 1 : GD-RST

GD-RST (uloc,pref[],A[])
  /* upos: query origin (latitude,ongitude) , pref[], an array consisting of user query preferences Q1,Q2,..Qn and A[], an array of their corresponding attributes a1,a2,..an */

1. Find the nearest vertex for the query origin uloc
        u ← NearestNeighborNode(uloc)
2. C[i]={0}  /*set the initial candidate objects as zero */
3. C[i] ←Generate(Q1)   /* retrieves the list of initial candidate set for the query point Q1*/
4. Heapset[i] ← addQPtoheap(C[])
5. if (checkskyline(Heapset[]) ) then
6.     Dispskylineobject()
            /* display  skyline object */
7.     Cacheskyline(C[], Heapset[])
8.         /* store for future queries*/
9.   end if

Each time a query is received, the dynamic skyline cache (DSC) stores the query point. If the cache is exceeded in size, the least frequently used (LFU) query is evicted. This is done by incrementing a counter to each skyline w.r.t query point and also considering the traffic update. The cache page will also be replaced when there is a drastic change in traffic.

### 4.3 Optimization

Optimization is achieved by computing the travel times to selected hotels which have the potential to participate in the final query answer instead of all hotels. For example, *(Query Q3):* Retrieve all low cost hospitals nearby restaurant and a temple. The optimization is as shown in Figure(4) and Figure(5).

Figure 4. Query Q3 before Optimization

| SNO. | POI_ID | LATITUDE | LONGITUDE | CATEGORY |
|---|---|---|---|---|
| 1 | H1 | 32.69139 | -117.28611 | LOW COST |
| 2 | H2 | 33.10917 | -117.325 | HIGH |
| 3 | H3 | 34.05333 | -118.18806 | MEDIUM |
| 4 | H4 | 33.72833 | -118.42667 | LOW COST |
| 5 | H5 | 34.17167 | -118.96889 | MEDIUM |
| 6 | H6 | 37.05528 | -118.61167 | LOWCOST |
| 7 | H7 | 33.15833 | -118.33667 | HIGH |
| 8 | H8 | 36.18222 | -118.40833 | MEDIUM |
| 9 | H9 | 34.15472 | -117.85944 | LOWCOST |
| 10 | H10 | 33.78 | -117.76056 | HIGH |

Figure 5. Query Q3 after Optimization

| SNO. | POI_ID | LATITUDE | LONGITUDE | CATEGORY |
|---|---|---|---|---|
| 1 | H1 | 32.69139 | -117.28611 | LOW COST |
| 2 | H4 | 33.72833 | -118.42667 | LOW COST |
| 3 | H6 | 37.05528 | -118.61167 | LOWCOST |
| 4 | H9 | 34.15472 | -117.85944 | LOWCOST |

From the above, *Query Q3* before and after optimization is compared. This approach would enhance the query processing by avoiding redundant extensive computations. Aggregations for secondary query points are done before applying skyline retrieval. This helps in further reduction of query computation.

## 5. EXPERIMENTAL EVALUATION

### 5.1 Experimental Setup

We conducted experiments on California road network which contains 21,050 nodes and 21693 edges. The algorithm is implemented in Java and tested on Windows Platform with Intel Core 2 CPU and 80GB memory. The main metric we adopt is CPU time that reflects how much time the algorithm takes in processing a skyline query. The input map is extracted from Tiger/Line files that are publicly available [23].

### 5.2 Results

#### 5.2.1 Impact of I/O on set of Query Points

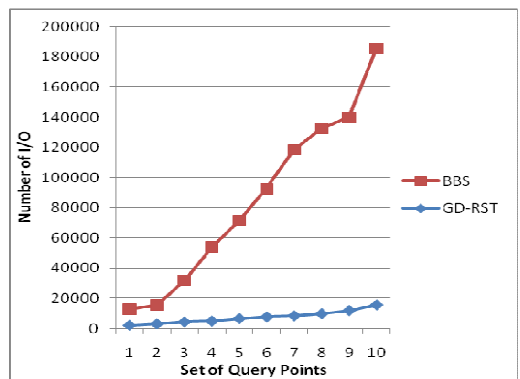

Figure 6: (a)

With this experiment in Fig 6(a), we show the impact of I/O on set of query points with GD-RST and Branch and bound Skyline (BSS) algorithms.

#### 5.2.2 Impact of CPU time on set of Query Points

Fig 6(b), we show the impact of CPU () on set of query points with GD-RST and BSS algorithms. Here total execution time of all query phases except pre-computation is taken for GD-RST and BBS algorithm is distance based skyline query processing. In both the comparison (fig 6(a) and 6(b)), the GD-RST algorithm has high performance than the traditional BSS algorithm. It has low I/O cost and less CPU time than its counterpart BSS.

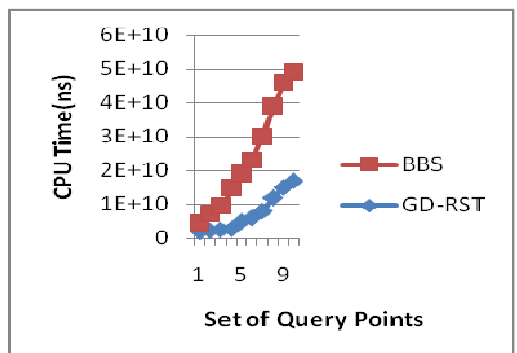

Figure 6: (b)

### 5.2.3 Impact of Cluster in search space

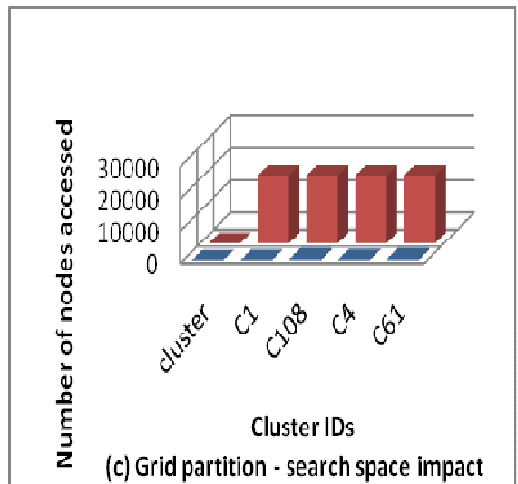

Figure 6: (c)

Fig 6(c), we show the impact of grid partitioning the road network to find nearest neighbor. Here C1,C108,C4,C61 are clusters that define the latitude, longitude boundary of an area. The impact of clustering with nodes in each of four cluster is depicted in figure 6(c).

### 5.2.4 Impact of cache in query computation

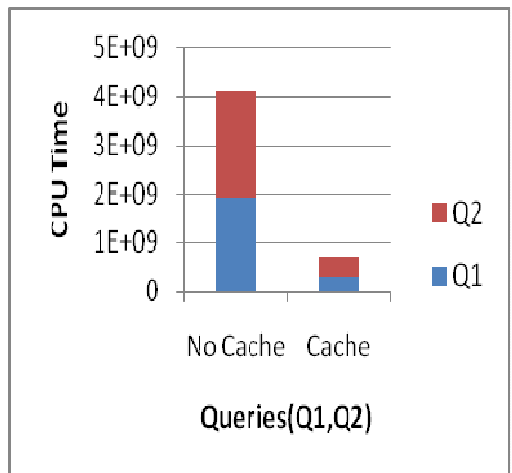

Figure 6: (d)

Fig 6(d), we show the impact of cache technique in computing future queries. Here Q1,Q2 are queries defined in section 3.2. The figure shows for processing future queries (i,e if same query is asked by same user/another user at any point of time and if query is in cache then the result of the query is taken from cache instead of computing it again. The CPU time reduction in using Dynamic Cache technique is shown here.

**5.2.5 Impact of query optimization on dominant objects**

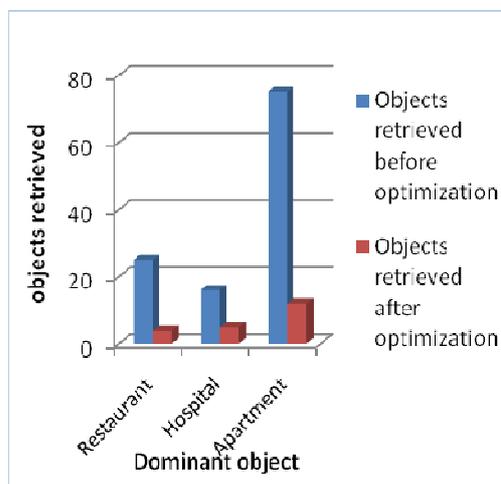

Figure 6: (e)

Fig 6(e), we show the impact of query optimization in object retrieval from database.

## 6. CONCLUSION

In this paper, we propose a Goal Directed Relative Skyline queries on travel time dependent (GD-RST) road networks. This algorithm efficiently searches and computes the nearest data object to query origin and in the user direction of traveling. The algorithm also caches the skylines to reduce the computation cost for future queries.

Additional future works includes skyline query for moving objects, different types of transportation mode, road types.